# Diagnostics of the structure of AGN's broad line regions with reverberation mapping data: confirmation of the two-component broad line region model

Ling Zhu[1] & Shuang Nan Zhang[2]†

[1] Department of Physics and Tsinghua Center for Astrophysics, Tsinghua University, Beijing 100084, China
[2] Key Laboratory of Particle Astrophysics, Institute of High Energy Physics, Chinese Academy of Sciences, Beijing 100049, China

**We re-examine the ten Reverberation Mapping (RM) sources with public data based on the two-component model of the Broad Line Region (BLR). In fitting their broad Hβ lines, six of them only need one Gaussian component, one of them has a double-peak profile, one has an irregular profile, and only two of them need two components, i.e., a Very Broad Gaussian Component (VBGC) and an Inter-Mediate Gaussian Component (IMGC). The Gaussian components are assumed to come from two distinct regions in the two-component model; they are Very Broad Line Region (VBLR) and Inter-Mediate Line region (IMLR). The two sources with a two-component profile are Mrk 509 and NGC 4051. The time lags of the two components of both sources satisfy $t_{IMLR}/t_{VBLR} = V^2_{VBLR}/V^2_{IMLR}$, where $t_{IMLR}$ and $t_{VBLR}$ are the lags of the two components while $V_{IMLR}$ and $V_{VBLR}$ represent the mean gas velocities of the two regions, supporting the two-component model of the BLR of Active Galactic Nuclei (AGN). The fact that most of these ten sources only have the VBGC confirms the assumption that RM mainly measures the radius of the VBLR; consequently, the radius obtained from the R-L relationship mainly represent the radius of VBLR. Moreover, NGC 4051, with a lag of about 5 days in the one component model, is an outlier on the R-L relationship as shown in Kaspi et al. (2005); however this problem disappears in our two-component model with lags of about 2 and 6 days for the VBGC and IMGC, respectively.**

## 1. Introduction

The Reverberation Mapping (RM) experiment has been used to diagnose the structure of the Broad Line Region (BLR) of Active Galactic Nuclei (AGNs) for decades. The broad emission lines are believed to be re-radiation of the central continuum. It is generally believed that the emission-line lightcurve responses to the continuum lightcurve through a "transfer function" which depends on the geometry and structure of the BLR[1]. The geometry and structure themselves cannot be determined through RM directly, neither can the transfer function. Cross correlation between the continuum lightcurve and emission-line lightcurve can obtain the lag of the emission-line lightcurve with respect to the continuum, and thus determine the average radius of the emission line region.

More information of the structure of the BLR comes from the profile of the emission line. A two-component model of BLR has been proposed based on the decomposition of broad Hβ lines[2-4]. This model suggests that the BLR is composed of two physically distinct regions, i.e., Very Broad Line Region (VBLR) and Inter-Mediate Line Region (IMLR). The broad Hβ line is the superposition of the emission lines from these two regions; they are Very Broad Gaussian Component (VBGC) and Inter-Mediate Gaussian Component (IMGC). The two regions have different radius and geometry and they response to the continuum variation individually; consequently, their emission-line lightcurves have different lags with respect to the continuum lightcurve. If this model is correct, we can do the following: (1) select the objects whose Hβ lines have double Gaussian profiles; (2) decompose them with a VBGC plus a IMGC; (3) obtain the lightcurves of these two component; (4) make cross correlations between each one and the continuum lightcurve, and; (5) ultimately obtain the lags of the two components with respect to the continuum, in order to determine the radii of the two emission regions (VBLR & IMLR) and thus test the two-component model.

In this paper we re-examine the RM data of ten sources in the public domain. This paper is organized in the following way: section 1 is an introduction; section 2 describes the line decomposition and basic properties of RM sources; section 3 is focused on the two sources that have double Gaussian profiles; section 4 shows a simulation to confirm the results of section 3; and section 5 is conclusion and discussion. Throughout this manuscript, "whole Hβ" means the whole broad Hβ line, after the narrow line is removed, "L5100" represents the intensity measured at 5100 Å and is used to indicate the continuum intensity.

## 2. Line decomposition and basic properties of the RM sources

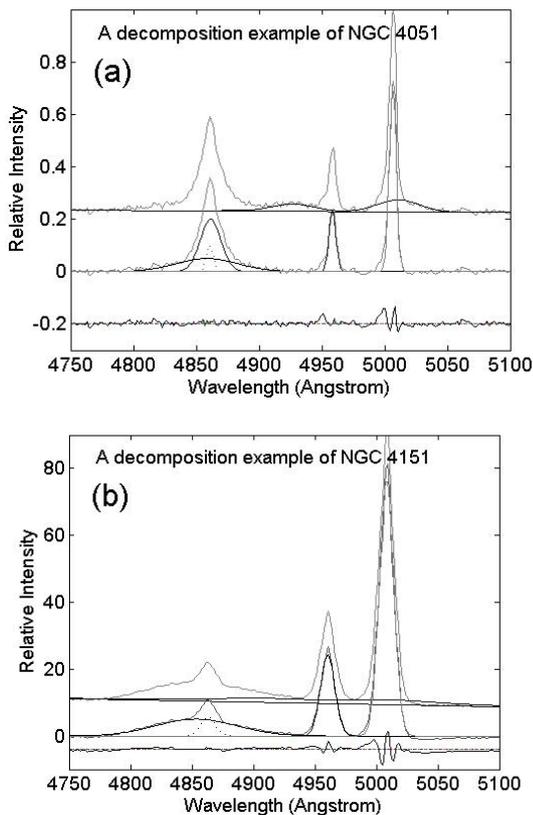

**Figure 1.** Decomposition examples. (a) is a spectrum with Hβ of a double Gaussian profile and (b) is a spectrum whose Hβ can be well described by one Gaussian component.

The spectroscopic monitoring experiments have lasted for many years and dozens of sources have been studied[5]. Ten of them have part of data public on AGN watch website (*http://www.astronomy.ohio-state.edu/agnwatch/*). The data for each source contain its continuum lightcurve, broad Hβ line lightcurve and a Hβ spectrum corresponding to each point in its lightcurve.

We decompose all the Hβ spectra of every source in the following way. Two examples are shown in figure 1; figure 1(a) is a decomposition example of NGC 4051 and figure 1(b) is an example of NGC 4151; the continuum is assumed to be a single power law. The Fe II feature is subtracted using the Fe II template[6]. The width of the narrow component of the Hβ is constrained to [OIII] double lines. Finally its broad component is fitted with a two-component model consisting of a VBGC and an IMGC. All the components are fitted at the same time using the least squares residuals fitting method. For Mrk 279, NGC4151,Fai 9, NGC7469, the average intensities of their IMGC are less than 20% of that of their VBGC. For Arkelian 564, the distribution of the FWHM of its IMGC is statistically indistinguishable from that of the narrow line component. For NGC 3783, the distribution of the FWHM of its IMGC is statistically indistinguishable from that of the VBGC. We therefore regard the above six sources to have only one broad Gaussian component, i.e., the VBGC. We then re-fit their spectra with just one broad component and the fittings are all statistically acceptable. In figure 1 (a) and (b) we show two examples for a two-component fitting and one-component fitting, respectively.

The fact that the broad Hβ of most of the sources (6 of 10) only need one Gaussian component to fit indicates that for most luminous sources the RM mainly measures the radius of the VBGC, as argued in Zhu, Zhang & Tang (2008)[4]. For the remaining four sources, one has a double-peaked profile (3C 390.3) and one has an irregular profile (NGC 5548). We do not investigate these above eight sources further in this work. Only two of them need two Gaussian components to fit; they are Mrk 509 and NGC 4051.

## 3. Detailed analysis of Mrk 509 and NGC 4051

### 3.1. Mrk 509

We acquire about five years' data of Mrk 509 from AGN watch website; they are the data used in Carone et al. (1996)[7]. All of their 151 Hβ spectra are decomposed as described in the last section. We calculate the flux of the IMGC and VBGC. As the spectra are not well calibrated, the intensities of IMGC and VBGC are calibrated by the intensity of [OIII] line (intensity of the narrow [OIII] line centred at 5007 Å is thought to be unchanged through the observation period). The lightcurves of different Gaussian components are shown in figure 2(b). Then we do cross correlation between the lightcurves of the whole broad Hβ and continuum. The data are not evenly distributed and thus linear interpolation is done to both lightcurves. The CCF centroid gives a lag of $82^{+9}_{-7}$ (all uncertainty regions correspond to $\pm 1\sigma$) days as shown in figure 2(c). This method is not exactly the same as usually used; the usually used interpolation method only interpolates one lightcurve and the other lightcurve is paired with this interpolated lightcurve[1]. However, our method gives nearly the same CCF as Carone et al. (2006)[7], who obtained a lag of about 80 days, within the uncertainty region of our result. The uncertainty for each centroid is obtained as described in Peterson et al. (1998)[8]. The comparison confirms that our method is also valid. Interpolating both lightcurves is easier to do and, more importantly, can make the comparison between real data and simulation directly, as shown in the next section. Cross correlation between VBGC (IMGC) and continuum lightcurve is done using the same method and a lag of $44^{+30}_{-5}$ ($138^{+5}_{-51}$) days is obtained. The CCFs are shown in figure 2(c). We can see that the CCF peaks of both VBGC and IMGC are not as significant as that of the whole broad Hβ; we will investigate this problem in the next section.

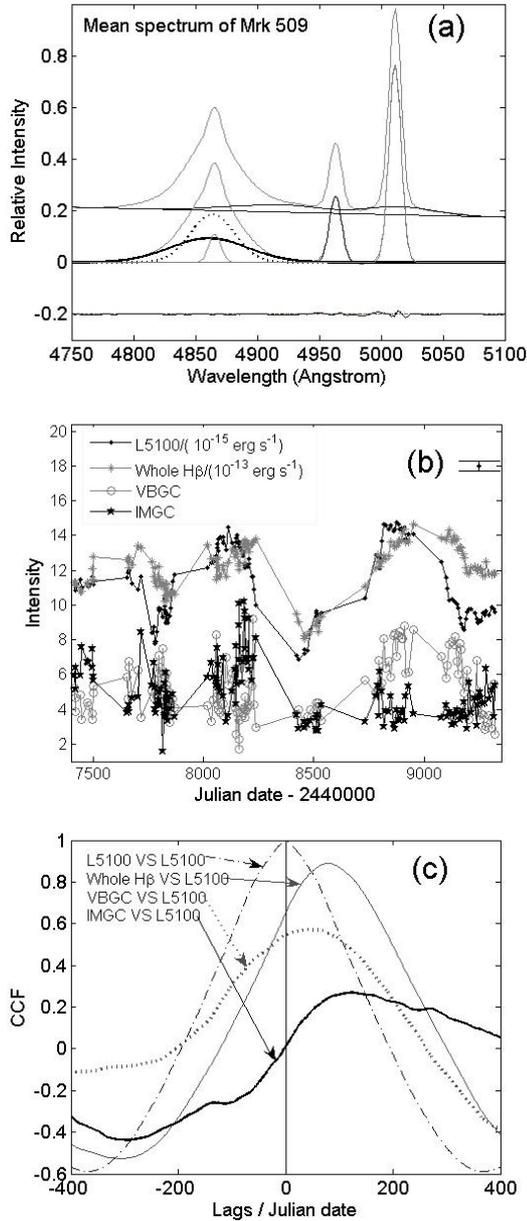

**Figure 2.** (a) is the decomposition of the mean spectrum of Mrk 509. The dashed line represents the IMGC, the thick black line represents the VBGC and gray lines represent narrow components. (b) shows lightcurves of Mrk 509. The intensity units of VBGC and IMGC are the same as whole Hβ. The typical errorbar is plotted on the top-right corner. (c) is Cross correlation Function of Mrk 509. The centroids of the CCFs are $82^{+9}_{-7}$, $44^{+30}_{-5}$ and $138^{+5}_{-51}$ days for the whole Hβ, VBGC and IMGC, respectively.

The mean spectrum can be well decomposed in the same way as we can see in figure 2(a); the FWHM of the IMGC is about 2560 km/s and that of VBGC is about 4610 km/s (the uncertainty of the FWHM of each Gaussian component is about 10%). According to Zhu, Zhang & Tang (2008)[4], if the two components are from two distinct regions whose motions are all dominated by the gravity of the same black hole, they should follow the relation $t_{IMLR}/t_{VBLR} = V^2_{VBLR}/V^2_{IMLR}$. Indeed Mrk 509 follows the relation as $138^{+5}_{-51} / 44^{+30}_{-5} \approx (4610/2560)^2$.

### 3.2. NGC 4051

The data in Peterson et al. (2000)[9] are public for NGC 4051, which is a Narrow Line Seyfert 1 galaxy (NLS1). There are about three years' data and 126 Hβ spectra. Only the lightcurve segment between Julian date 2450180 to 2450260 is used following Peterson et al (2000)[9]. The same process is done to NGC 4051 as to Mrk 509. As shown in figure 3, the mean spectrum of broad Hβ can be well described by the two-component model with FWHM of 3080 km/s and 1240 km/s, respectively. The CCF of the whole Hβ lightcurve versus continuum lightcurve is quite similar to that in Peterson et al. (2000)[9]. The centroids of CCF of VBGC and IMGC lightcurves versus continuum lightcurve are about $2_{-1.6}^{+1.7}$ and $6_{-1.5}^{+2.2}$ days, respectively. It also follows the relation $t_{\text{IMLR}}/t_{\text{VBLR}} = V_{\text{VBLR}}^2/V_{\text{IMLR}}^2$ as $6_{-1.5}^{+2.2}/2_{-1.6}^{+1.7} \approx (3080/1240)^2$. Similar to Mrk 509, the CCF peaks of VBGC and IMGC versus the continuum is not as significant as that of the whole broad Hβ versus continuum.

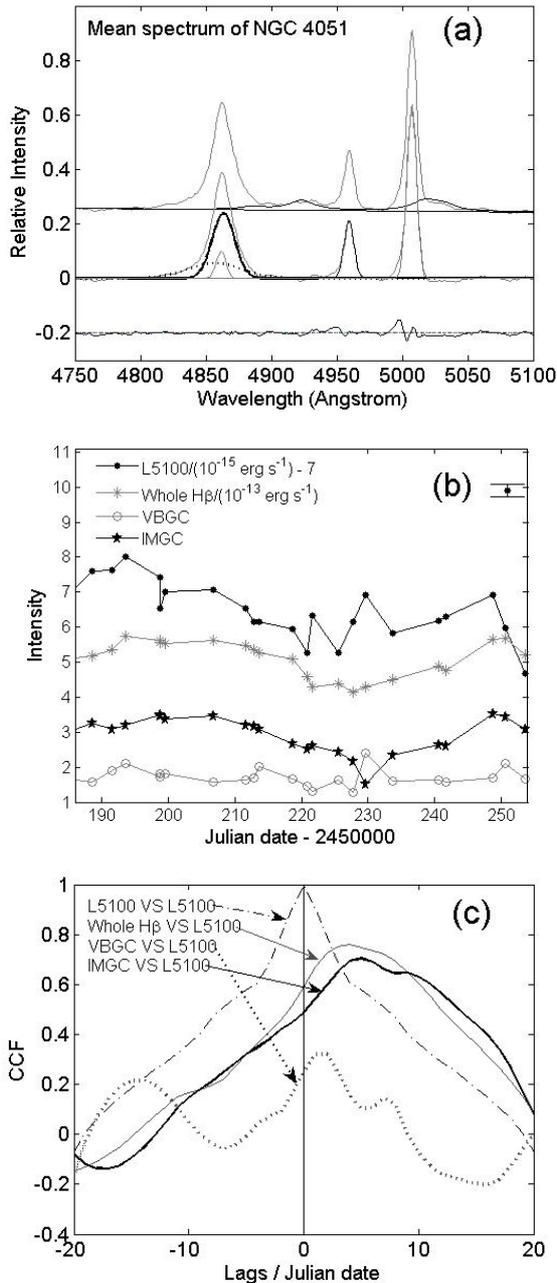

**Figure 3.** (a) is decomposition of the mean spectrum of NGC 4051. The dashed line represents the IMGC, the thick black line represents the VBGC and gray lines represent narrow components. (b) is Lightcurves of NGC 4051, the intensity units of VBGC and IMGC are the same as whole Hβ. The typical errorbar is plotted on the top-right corner. (c) is cross correlation Function of NGC 4051. The centroids of the CCFs are $5_{-1.1}^{+1.1}$, $2_{-1.6}^{+1.7}$ and $6_{-1.5}^{+2.2}$ days for the whole Hβ, VBGC and IMGC, respectively.

NGC 4051 is an outlier of the BLR radius-continuum luminosity (R-L) relationship as shown in Kaspi et al. (2005)[5]. Its radius of $5_{-1.1}^{+1.1}$ lt-days obtained from cross correlation between the whole broad Hβ and the continuum lightcurve is too large for it to follow the R-L relationship[5]; the R-L relationship favours this object to have a radius of about one lt-day as we can see in figure 4. One lt-day is just in the uncertainty range of the lag between VBGC and continuum lightcurve which represents the radius of VBLR in the two-component model. Mrk 509 is also marked on figure 4, showing a better consistency with the R-L relationship, althought it was not an obvious outlier of the R-L relationship previously. This is additional evidence that the R-L relationship obtained from RM may mainly reflect the behaviour of the VBLR.

## 4. Simulation

A main problem of the above results is that the CCF peaks of both VBGC and IMGC versus the continuum lightcurve are not as significant as that of the whole Hβ lightcurve. If the decomposition of the two-componet model is physical rather than a mathematical description, the meaning of intensity of a single Gaussian component will be more definite, and thus the CCF peak of a single Gaussian component versus the continuum should be more significant. However, the whole broad Hβ intensity is from the direct observation and the VBGC or IMGC intensity comes from decomposition which may bring some additional uncertainties due to the decomposition process, as confirmed by our simulations shown as follows.

The simulation is done to the two sources in the same way and thus we will only describe it using Mrk 509. The continuum lightcurve of Mrk 509 after interpolation in section 3 is used as the original lightcurve of the source, and the lightcurves with 44 and 138 days' delay of the original lightcurve are used as the VBGC and IMGC lightcurves, respectively. A spectrum at every point of the lightcurve is then created. Every spectrum is a superposition of a VBGC, an IMGC, a narrow Hβ line, double [O III] lines, Fe II lines and noise; the noise is taken as the rms of the residuals of the real spectral decomposition. The widths of the VBGC and IMGC are fixed at the value of the mean spectrum (4610 and 2560 km/s) and the heights of them vary with their own lightcurves. The lightcurve of the VBGC has been multiplied by a constant number of 0.59 to ensure that the average intensity ratio of the VBGC to IMGC is the same as to the real sample in section 3. Other components are all created with parameters the same as to the mean value of the real sample of Mrk 509; a random variation with a range of 20% is added to the width and a random variation with a range of ±5Å is added to the central position of these Gaussian components. A series of spectra are created in such a way that the broad Hβ is truly a superposition of two Gaussian components with different lags to the continuum. We then decompose these simulated spectra and obtain the intensities of the VBGC, IMGC and whole Hβ. New lightcurves of the VBGC, IMGC and whole Hβ line are thus produced. We do cross correlation between these lightcurves and the original lightcurve of the continuum. Figure 5(a) and 5(b) show the similar simulation results for NGC 4051 and Mark 509.

Three things are important in figure 5(a). First, the lags of VBGC and IMGC to the continuum are 46 and 130 days, respectively, nearly the same as the input values. This indicates that the intrinsic lags can be recovered properly with our decomposition and cross correlation method. Second, the lag of the whole broad Hβ to the continuum is 84 days which is nearly the same as that we obtained from the real sample. The superposition of the two components really re-produces the whole Hβ lightcurve as observed. Third, The CCF peaks of IMGC and VBGC are also not as significant as that of the whole Hβ versus the continuum lightcurve, proving that the decomposition process can indeed decrease the peak values of CCF of both the IMGC and VBGC, due to the limited signal to noise ratio of the data.

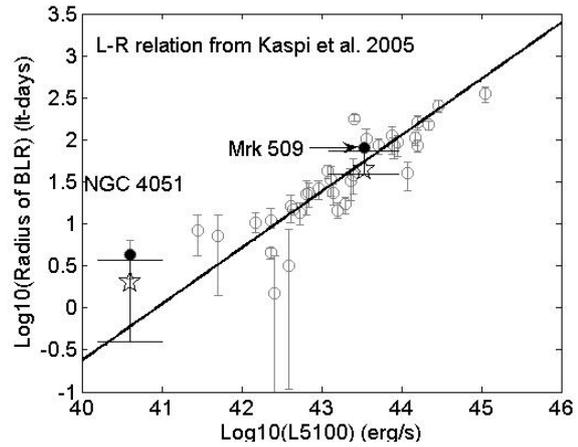

**Figure 4.** The L-R relationship from Kaspi et al 2005; NGC 4051 lies upon the relationship. The solid line is the best fitting. For NGC 4051, the new radius $2^{+1.7}_{-1.6}$ lt-days of its VBLR in the two-component model makes it consistent with the L-R relationship. For Mrk 509, its VBLR radius also matches the R-L relationship better.

Although our simulation results show some similarities with the real data, the profiles of the simulated CCF are not exactly the same as the CCF obtained from the real data. This may be due to the complicated re-radiation processes which produce the broad lines, because the line emission regions do not always follow the continuum flux variation in a linear fashion; however a linear response is assumed in our simulation for simplicity. Nevertheless our simulation study suggests that the weaker CCF of the two components versus the continuum, compared to that of the whole line, does not invalidate our model that the two Gaussian components come from physically two distinct regions.

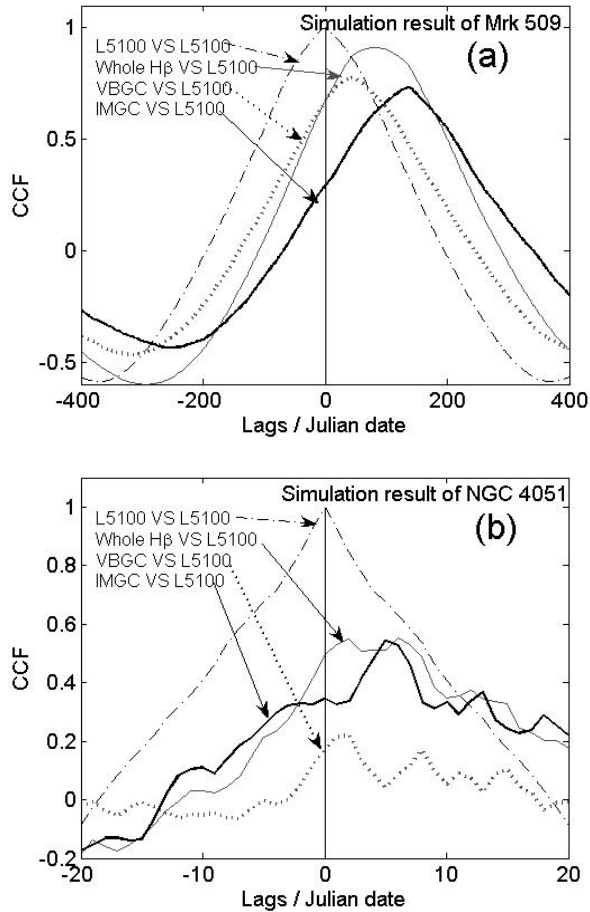

**Figure 5.** (a) is the simulation result for Mrk 509; the input lags are 44 and 138 days for the VBGC and IMGC, respectively. The CCF centroids of this figure are 84, 46, and 130 for the whole Hβ, VBGC and IMGC, respectively. (b) is the simulation result for NGC 4051; the input lags are 1 and 6 days for the VBGC and IMGC, respectively. The CCF centroids of this figure are 4, 2, and 5 for the whole Hβ, VBGC and IMGC, respectively.

## 5. Conclusion and discussion

RM data of ten sources are re-examined. In fitting their broad Hβ lines, six of them only need one Gaussian component, and two of them, namely, Mrk 509 and NGC 4051, need two Gaussian components. Two different time lags are obtained when cross correlations of VBGC and IMGC versus the continuum lightcurve are done, for each of the two sources. The two time lags and the widths of the two Gaussian components follow the relation $t_{IMLR}/t_{VBLR} = V_{VBLR}^2/V_{IMLR}^2$, which can be naturally explained in the two-component model of the BLR. The result supports that the broad Hβ line is a superposition of two Gaussian components which come from two distinct emission regions. The two emission regions have different radii and they respond to the continuum variation individually.

As most of the RM sources have broad Hβ lines with a single Gaussian profile, RM mainly measures the radius of the VBLR, thus the R-L relationship based on RM mainly reflects the behaviours of VBLR. NGC 4051 follows the R-L relationship if the lag obtained from cross correlation between the VBGC, instead of the whole broad Hβ, and the continuum lightcurve is used. It also supports that R-L relationship represents radius of VBLR.

The radius obtained from RM bears certain uncertainties because the response of the emission region to the continuum variation is not always the same. Therefore the radii obtained from different observation segments may have significant differences. In the four subsets of RM data of Mrk 79, only one peak emerges at about 10 days in the CCCD map of the first three subsets. However, another peak at about 40 days appears in the CCCD of the fourth subset[10]. The 10 and 40 light days are explained as the radii of the VBLR and IMLR in Zhu, Zhang & Tang (2008). Recently, new observations of NGC 4051 gives a lag of the whole Hβ with respect to the continuum of about 1.75 days[11] which is near the $2^{+1.7}_{-1.6}$ days lag between the VBGC and the continuum in the observations we analyzed here. In our analysis based on our two-component model, the lags obtained from the whole Hβ are always between the lags obtained from the VBGC and IMGC. Our simulation shows that it will be much nearer to that obtained from VBGC when the broad Hβ intensity is dominated by VBGC; the same is also true for the IMGC. Since the profile of the whole Hβ varies from time to time, it seems that the lags from the whole Hβ may mainly reflect the radius of VBLR at some time and IMLR at other times, if the physical conditions of the two emission regions change between different periods of observations separated by months or years. This may provide an explanation to different lags of the whole Hβ obtained at different observation periods.

For RM experiments focused on relatively low luminosity AGNs like NGC 4051[11], a large percentage of these AGNs are NLS1s whose broad Hβ are dominated by the IMGC[4]. The lag obtained from the whole Hβ may be near the radius of the IMLR and thus significantly different from that of the VBLR. If the previously established R-L relationship mainly reflects radius of VBLR, we predict that the lags of whole Hβ of these NLS1s will deviate the R-L relationship, but the lags obtained from VBGC should continue to follow the relationship. We further predict that NLS1s should follow this new R-L relationship for the IMGC,

$$\log\left(\frac{R_{\text{IMLR}}}{\text{cm}}\right) = 1.1^{+2.5}_{-2.5} + 0.37^{+0.06}_{-0.06} \log\left(\frac{L5100}{\text{erg s}^{-1}}\right),$$

which was established in our previous work based SDSS data[4]

*SNZ acknowledges partial funding support by Directional Research Project of the Chinese Academy of Sciences under project No. KJCX2-YW-T03 and by the National Natural Science Foundation of China under grant Nos. 10821061, 10733010,10725313, and by 973 Program of China under grant 2009CB824800.*